УДК 004.9 + 520.84

# Индекс каталог публикаций «Известий Крымской астрофизической обсерватории»

*К.В. Логачев[1], А.А. Шляпников[2]*

[1] ФГБУН «Крымская астрофизическая обсерватория РАН», Научный, Крым, Россия, 98409
*kir@craocrimea.ru*
[2] ФГБУН «Крымская астрофизическая обсерватория РАН», Научный, Крым, Россия, 98409
*aas@craocrimea.ru*

**Аннотация.** Представлена информация о создании «Индекс каталога» для статей, опубликованных в журнале «Известия Крымской астрофизической обсерватории». Описана краткая история накопления ссылок для каталога, его структура и схема взаимодействия с основными мировыми астрономическими базами данных.
 INDEX CATALOGUE FOR THE ``IZVESTIYA KRYMSKOJ ASTROFIZICHESKOJ OBSERVATORII'' PUBLICATIONS, by K.V. Logachev, A.A. Shlyapnikov. Information on the creation of the index-catalogue for articles published in the journal "Izvestia of the Crimean Astrophysical Observatory" ("Izvestiya Krymskoj Astrofizicheskoj Observatorii" from ADS Bibliographic Codes: Journal Abbreviation - IzKry) is presented. A brief history of the accumulation of references for the catalog, its structure and the scheme of interaction with the basic world astronomical databases are described.

**Ключевые слова:** каталог, публикации, базы данных

## 1 Введение

Индекс-каталог (далее – Каталог) является основным элементом базы данных (БД) публикаций Крымской астрономической виртуальной обсерватории (Шляпников, 2015). В него входит информация об объектах, наблюдавшихся сотрудниками в Крымской астрофизической обсерватории или других организациях, данные для которых были опубликованы в первых 100 томах «Известий Крымской астрофизической обсерватории» (далее – «Известия»). Последующие тома формировались в машиночитаемых форматах, что упрощает контекстный поиск информации в них.

 Необходимость создания такого Каталога обусловлена неполным представлением в мировых астрономических базах данных сведений о публикациях в «Известиях», и ещё меньшим указанием конкретных объектов, описанных в статьях. Для создания Каталога был составлен независимый список всех публикаций по первым 100 томам «Известий». В наш список были добавлены стандартные библиографические коды публикаций согласно классификатору

SAO/NASA Astrophysical Data System (далее – ADS). На основе этой информации был произведен анализ представления в ADS статей «Известий» и ссылок на объекты в базе данных SIMBAD. Более подробно проблема наполнения ADS информацией о публикациях в «Известиях» рассмотрена в статье (Шляпников, 2007, Бондарь, 2013).

## 2 Описание структуры каталога

В первом приближении Каталога информация об объектах бралась из названия статей, во втором – из их содержания. Отдельными списками для включения в Каталог представлялись объекты, описанные в статьях, посвященных каталогизации, где речь шла о наблюдениях десятков, сотен и тысяч объектов. Прототипом Каталога стала статья «Небо пятидесятки»… (Шляпников, 2013), в которой описан каталог и библиография объектов, наблюдавшихся на 50″ телескопе Крымской астрофизической обсерватории.

Структура Каталога включает одиннадцать колонок:
– координаты объекта на эпоху 2000.0 года по SIMBAD;
– базовое обозначение объекта в SIMBAD;
– обозначение объекта в статье «Известий»;
– тип объекта по SIMBAD;
– звёздная величина по SIMBAD;
– спектральный класс объекта по SIMBAD;
– время наблюдений (начало и конец);
– инструмент, на котором проводились наблюдения;
– краткое описание (ключевые слова) представленных данных;
– библиографический код ADS;
– ссылка MPC (M), NED (N), VizieR (V) если она связана с публикацией в «Известиях».

Пример фрагмента таблицы Каталога представлен на рисунке 1.

| № | R.A.(2000.0) | Decl.(2000.0) | SIMBAD | IzKry | Type | Vmag | Spectr | ADS Bib.code |
|---|---|---|---|---|---|---|---|---|
| 1 | 00 00 43.63438 | +45 15 11.9987 | V* CG And | HD 224801 | a2* | 6.305 | B9p… | 1971IzKry..43..113L |
| 2 | 00 01 38.6309 | +60 26 59.717 | HD 224905 | HD 224905 | Be* | 8.58 | B1Vn | 1969IzKry..39...63V |
| 3 | 00 01 39.46002 | +73 36 42.6415 | HD 224890 | HD 224890 | * | 6.687 | Am… | 1980IzKry..62...34D |
| 4 | 00 02 10.1552 | +27 04 56.122 | GJ 914 A | 85 Peg | *i* | 6.41 | ~ | 1958IzKry..18....3M |
| 70 | 00 06 03.38745 | +63 40 46.7605 | HD 108 | HD 108 | SB* | 7.494 | O4-8f?p | 1964IzKry..32..108G |
| 71 | 00 06 03.38745 | +63 40 46.7605 | HD 108 | HD 108 | SB* | 7.494 | O4-8f?p | 1965IzKry..33..242K |
| 80 | 00 06 36.78482 | +29 01 17.4038 | V* V439 And | BD+28 4704 | BY* | 6.88 | K0V | 1958IzKry..18....3M |
| 111 | 00 08 23.25988 | +29 05 25.5520 | V* alf And | HD 358 | a2* | 2.012 | B9II | 1960IzKry..22..225M |
| 112 | 00 08 23.25988 | +29 05 25.5520 | V* alf And | HD 358 | a2* | 2.012 | B9II | 1960IzKry..23..148K |
| 115 | 00 08 23.25988 | +29 05 25.5520 | V* alf And | HD 358 | a2* | 2.012 | B9II | 1978IzKry..58...81B |
| 121 | 00 09 10.68518 | +59 08 59.2120 | V* bet Cas | HD 432 | dS* | 2.61 | F2IV | 1954IzKry..12..148M |
| 122 | 00 09 10.68518 | +59 08 59.2120 | V* bet Cas | HD 432 | dS* | 2.61 | F2IV | 1958IzKry..18....3M |
| 123 | 00 09 10.68518 | +59 08 59.2120 | V* bet Cas | HD 432 | dS* | 2.61 | F2IV | 1960IzKry..23..148K |
| 131 | 00 13 14.15123 | +15 11 00.9368 | V* gam Peg | HD 886 | bC* | 2.60 | B2IV | 1958IzKry..20..123K |
| 143 | 00 17 05.49885 | +38 40 53.8902 | * tet And | HD 1280 | ** | 4.679 | A2V | 1960IzKry..23..148K |
| 155 | 00 17 43.06183 | +51 25 59.1242 | V* AO Cas | AO Cas | El* | 5.955 | O9IIInn+… | 1967IzKry..37..205G |

Рис. 1. Пример фрагмента Каталога в структурированном текстовом формате.

## 3 Структура каталога в VOTable формате

Каталог подготовлен в форматах, поддерживаемых приложениями виртуальной обсерватории, и обеспечен гиперссылками к базам данных ADS, MPC, NED, SIMBAD и VizieR. Его структура в VOTable формате соответствует основной структуре описанной ранее, представленной в XML разметке. Это обеспечивает визуализацию Каталога в интерактивном атласе неба Aladin (рис. 2) и других приложениях Международной виртуальной обсерватории и переход к указанным выше базам данных.

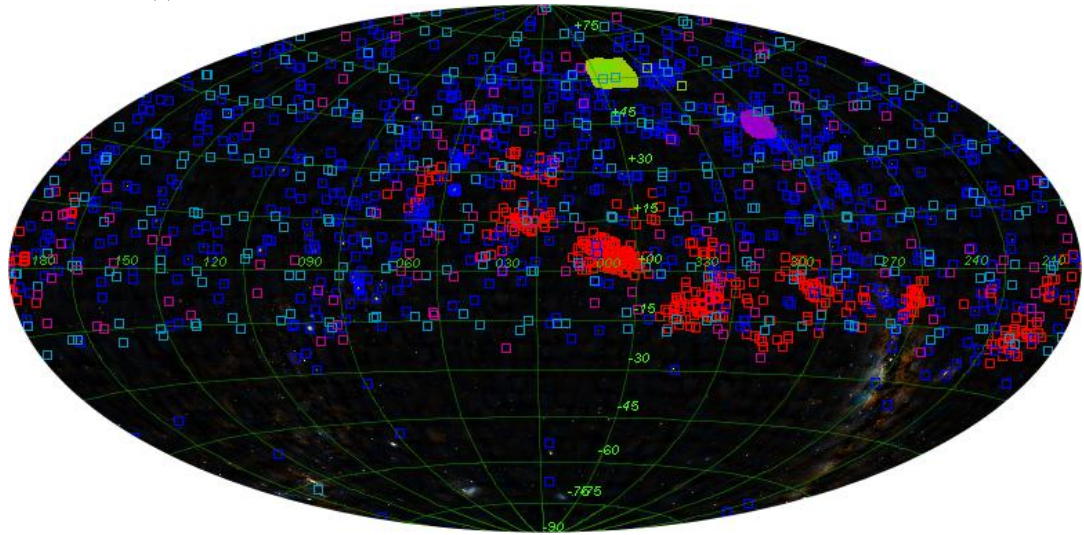

Рис. 2. Представление Каталога в атласе неба Aladin (проекция AITOFF). Различными цветами показаны библиографические ссылки из «Известий», полученные на разных инструментах.

Наличие в Каталоге гиперссылки к БД SIMBAD обеспечивает переход к наиболее полной информации об объекте, содержащейся в этой базе данных . Включение в Каталог библиографического кода SAO/NASA ADS позволяет перейти по соответствующей ссылке к информации о статье из «Известий», размещённой на сервисе астрономических абстрактов.

В первых томах «Известий» публиковались статьи с результатами астрометрических измерений для малых тел Солнечной системы. С целью доступа к информации о наблюдавшихся объектах Каталог содержит ссылку на сайт IAU Minor Planet Center – MPC.

Для объектов, содержащихся в БД NASA/IPAC Extragalactic Database – NED в VOTable предусмотрена гиперссылка, также как и для объектов/каталогов в БД VizieR.

## 4 Заключение

При создании Каталога активно использовались поддерживаемые Центром астрономических данных в Страсбурге приложения SIMBAD, VizieR и ALADIN, библиографический сервис SAO/NASA ADS, информация Центра малых планет Международного астрономического союза (IAU MPC) и база данных внегалактических объектов NASA/IPAC (NED). Авторы признательны всем, кто обеспечивает их работу.

## 5 Литература

Индекс каталог публикаций «Известий Крымской астрофизической обсерватории»

Шляпников и др. (Shlyapnikov A., Bondar' N. and Gorbunov M.) // Baltic Astronomy. 2015. V. 24. P.462.